# Title: Universal coupling between the photonics and phononics in a 3D graphene sponge

M. Shalaby[1,2,†], C. Vicario[3], F. Giorgianni[3,4], M. A. Gaspar[3], P. Craievich[3], Y. Chen[5], B. Kan[5], S. Lupi[6] and C. P. Hauri[3]

- [1]Beijing Advanced Innovation Center for Imaging Technology and Key Laboratory of Terahertz Optoelectronics, CNU, Beijing 100048, China
- [2]Swiss Terahertz Research-Zurich, Technopark, 8005 Zurich, Switzerland and Park Innovaare, 5234 Villigen, Switzerland
- [3]Paul Scherrer Institute, 5232 Villigen, Switzerland
- [4]Center for Life Nano Science@Sapienza, Istituto Italiano di Tecnologia, V.le Regina Elena 291, I-00186, Roma, Italy; and Department of Physics, Sapienza University of Rome, Piazzale Aldo Moro 2, I-00185 Roma, Italy.
- [5]State Key Laboratory and Institute of Elementog-Organic Chemistry, Nankai University, Tianjin 300071, China
- [6]Graphene Labs, Istituto Italiano di Tecnologia, Via Morego 30, 16163 Genova, Italy

[†] *Corresponding authors:* <u>most.shalaby@gmail.com</u>

**Abstract**: Photon-phonon coupling holds strong potential for sound and temperature control with light, opening new horizons in detector technology, remote sound generation and signal broadcasting. Here, we report on a novel stereoscopic ultralight converter based on a three dimensional graphene structure 3G-sponge, which exhibits very high absorption, near-to-air density, low inertia, and negligible effective heat capacity. We studied the heat and sound generation under the excitation of electromagnetic waves. 3G-sponge shows exceptional photon to heat and sound transduction efficiency over an enormous frequency range from MHz to PHz. As an application, we present an audio receiver based on a 3G-sponge amplitude demodulation. Our results will lead to a wide range of applications from light-controlled sound sources to broadband high-frequency graphene electronics.

**One Sentence Summary:** We present a universal photon-phonon coupling in a 3D graphene sponge for efficient heat and sound emission.

**Main Text:**

**Introduction:** Photons and phonons are both energy quanta that can be associated with waves oscillating in time and space at various frequencies. However, their generation, propagation, and interaction with matter are fundamentally different, respectively constituting photonic (electromagnetic, EM) and phononic spectra. While the phonons carry sound through mechanical vibrations and require the presence of a medium, the EM waves propagate even in vacuum. Photon-phonon coupling control is an exciting but rather an unexplored route to manipulate heat and sound emission with light. The process involves the physical transfer of energy through vibrations. This makes sound waves propagate much further than heat waves. Light-controlled phononics is foreseen to open up a broad range of novel applications including electromagnetic detection, remote sound generation technologies and broadband signal processing [1].

The efficiency of energy transfer from EM waves to sound and heat (photon-phonon coupling) is the present main obstacle for these applications. Sound and heat can be coupled and the high frequency heat energy (THz range) can be converted into sound energy (kHz range). Such coupling is complex. Elasticity and inertia are the most relevant medium properties for sound and mechanical wave generation.

Here we present the 3D graphene sponge (3G-sponge) as a distinguished material for a universal coupling between photons and phonons. 3G-sponge is a newly-developed material [2, 3] that shows intriguing properties such as super-compressive elasticity and near zero Poisson's ratio [4]. These properties are prerequisites for a highly efficient photon-phonon coupling [5]. Recent studies of mono-layer (i.e. 2D) graphene have reported the outstanding molecular properties in view of thermal conductivity [6], charge carrier mobility and Young' modulus [7, 8, 9]. But, so far, little attention has been paid to the equivalent macroscopic properties of 3D graphene, as stacking graphene sheets to form a bulk has always compromised the interesting properties of 2D graphene. It is only very recently that Wu et al. succeeded in developing the graphene foam [4], a bulk 3D free-standing graphene sponge structure - the lightest solid material on earth.

**Results:** By exploiting its unique properties, we present the 3G-sponge as a universal coupler between light, sound and heat with exceptional efficiency. The proposed concept of photon-phonon coupling is shown in Fig. 1. We explored two different excitation schemes of the 3G-sponge with a time-continuous EM wave (CW) and amplitude-modulated (AM) stimuli. Due to the high absorption capability [10] and the negligible heat capacity of the 3G-sponge [3-4] the excitation with a CW wave leads to a rapid rise in temperature.

To prove this fundamental coupling concept experimentally, we started our investigation with the CW microwave radiation source shown in Fig. 1A. To reduce free-space losses the 3G-sponge is directly attached to the coaxial cable and is irradiated by an EM sine wave with an average power of up to 1 W and GHz frequency. Upon EM excitation, we observed a fast rise of the 3G-sponge temperature, which rapidly reaches a steady state. The final temperature was studied for different excitation wave frequencies (0.5- 4 GHz, Fig. 1B). We found linear dependence on the average power that the rise in temperature increases as the excitation frequency decreases (Fig. 1B). For the lowest excitation frequency (0.5 GHz) the 3G-sponge heats up beyond 150°C, which represents the maximum temperature limit of our thermographic camera. The fast rise in temperature originates from the remarkable absorption properties of the 3G-sponge to the used microwave [10].

While the CW stimulus gives rise to extensive heat emission of the 3D-sponge, no emission of sound could be observed. The physical scenario changed when the CW stimulus was modulated in intensity. For this, the CW GHz source was modulated in amplitude with a 1-50 kHz sine wave (experimental setup shown in Fig. 1C). The sponge illuminated with such an AM stimulus reproduces a thermal wave modulated at the same frequency. However, direct recording of such a fast heat transient is technically challenging. Instead, we experimentally observed an audible acoustic signal from the sponge caused by the modulated heat wave. Indeed, using a conventional microphone, we could measure a powerful acoustic signal emitted from the 3G-sponge at precisely 5 kHz frequency (Fig. 1D), in correspondence to the 5 kHz EM stimulus modulation frequency. This surprising result demonstrates that the sound generation is directly linked to photon absorption, leading to temperature transients on a fast, microsecond timescale. The corresponding emitted sound wave is shown in black in Fig. 1D) together with the modulating signal (blue). Sound emission occurs exactly at a $\pi/2$ phase shift with respect to the

EM stimulus. This is in line with the expectations that the level of sound is a function of the derivative of the EM intensity signal. From these data, it is evident that the emitted sound wave follows the temporal dynamics of the modulated EM stimulus. It is worth mentioning that this efficient coupling mechanism enabled loud sound generation which was easily perceptible by the human ear.

Graphene-based materials are known for their extremely wide spectral absorption [11]. Therefore, we expanded our study on sound emission and heating from the microwave (GHz) to the terahertz (THz) [12, 13] and optical (PHz) frequency ranges. For this, we used a pulsed laser stimulus. The pulsed THz source (see Material and Methods section) provides a sub-picosecond single-cycle pulse with main spectral contents located in the 1-12 THz range at a repetition rate of 100 Hz [12]. To study spectrally- dependent response, we used different THz low pass filters (LPFs) with cut-off frequencies at 2, 3, 4.2, 6, 10 and 20 THz, respectively. In this measurement, the 3G-sponge is placed at the focus of the THz beam [12]. The audio signal level scale linearly with the THz pulse energy and it is independent of the THz spectrum. The microphone output signals obtained with different THz spectra clearly show the pulsed time structure of the THz source (Fig. 2A lower graph). Similar emission of sound and heat is also triggered by a pulsed femtosecond laser source operating at carrier frequencies in the optical regime ($\nu_{ph}$=200 THz, $E_{ph}$=0.83 eV) as shown in Fig. 2B and by a microwave source (Fig. 2C). The thermal image of the 3G-sponge irradiated by the THz beam , Fig. 2E, shows that the heat emission and consequently the sound generation are localized within the sub-mm THz focus where the temperature increases by 3.1 °C. Similar to the microwave measurements, we observed a linear dependence of the sound signal and heating on the THz and PHz pulse energy impinging on the 3G-sponge.

A summary of the acoustic spectral output for different stimuli is given in Fig. 2D. Independent of the laser stimulus central frequency, our experiment shows the emission of broadband sound spectra (0.1-50 kHz) with minor qualitative differences. This suggests that the coupling between the EM waves to sound in the 3D-sponge is nearly independent of the pulsed stimulus central frequency. This establishes graphene sponge as a universal coupling medium between phonons and photons across an extremely large frequency range from the far infrared to the visible. It allows for efficient heat and sound generation by simple illumination with an amplitude-modulated light beam. While the reported phenomenon of sound emission relies on transient photon absorption and heating, we stress that the sound generation mechanism observed in our experiments depends mainly on the pulse energy rather than the peak power. Therefore, a high repetition rate pulsed laser or a modulated continuous wave THz source are better suited for sound emission applications.

**Discussion:** The physical origin of the ultra-broadband coupling mechanism depends on the stimulus' center frequency. 3G-sponge can be considered as interlaced carbon sheets. Optical phonons in graphene occur at high frequency (~6 μm, 50 THz) [11,14-17]. This optical phonon which is not IR active in single layer carbon (not dipole active at the Gamma point) turns active in bilayer and multilayered carbon (shows net dipole at the Gamma point). Due to the high excitation frequency there is no direct path of excitation of such a phonon with a microwave and far infrared stimulus. The photons in this low-frequency THz range are rather absorbed by Dirac fermions, i.e. high mobility electrons. After absorption the electron-phonon scattering process during which the energy gained by electrons is transferred rapidly to phonons takes place on the order of 10-100 ps [15]. In graphene, this electron-phonon scattering process is efficient. It was recently shown that at high THz electric field, significant fraction of carriers can go beyond the (Dirac point) phonon energy (0.2 eV) [15]. This leads to a corresponding increase in the optical phonon population which consequently results in a reduction in electron mobility, THz absorption, and sound generation [17]. For a stimulus at optical and near infrared frequency, the 3G-sponge still shows perfect absorption but the underlying energy transfer mechanism is different. At photon energies $E_{\text{ph}} > 2\,\mu_e = 400\,me\text{V}$ ($\mu_e$: the chemical potential) electronic interband transitions are excited. The absorbed energy is dissipated by the relaxation of the excited carriers and emission of optical phonons [16, 17]. The photon-electron absorption in graphene depends on the Fermi energy. For a doped system, one finds absorption from free Dirac electrons and absorption from interband transitions which is already happening at $\omega > 2\mu_e$ (Pauli blocking). In 3G-sponge, one expects that $E_F$ is nearly zero, implying a strong reduction of free electron absorption and interband transitions already happening around $\omega$ close to DC. In this regard, while $2\mu_e = 400$ meV seems to be a good approximation, its exact value depends on the specific doping of graphene. We stress that the above-mentioned picture is based on the studies of doped graphene layers. 3G-sponge is a complex disordered insulator which implies modified and perhaps different thermodynamics. Yet, the most probable interpretation is that the microwave absorption depends mainly on the free charge carriers already present in the systems [4]. The photon-phonon interaction at higher frequencies (such as n-IR) is mediated by electrons. During the sound generation, the thermal barrier is not expected to play a major role as the sound generation occurs on a much slower time scale (microsecond) than the heating (picoseconds).

Despite the difference in the underlying photon absorption mechanisms in 3G-sponge below and above the 3G-sponge bandgap our measurements confirm the universality of the heat and sound generation across the extremely large electromagnetic frequency range from GHz to PHz. We note that carbon nanotubes have been used in the past for ultrasound generation and that graphene sheets have been demonstrated as thermo-acoustic transducers [18]. The advance of our work is based on the unique physical properties of the 3G-sponge structure, which enables universal absorption and coupling of the EM radiation to phonon and sound generation with unprecedented high efficiency. From the application point of view, the characteristic time scale of light conversion into sound is an important parameter. First, light-heat conversion is mediated by electron-phonon scattering occurring on the sub-picosecond time scale [15-17]. Second, heat-sound generation occurs through modulation of the air pressure inside the sponge taking place. In our measurements, the maximum sound transient time was around 10 microseconds limited by the spectral response of our microphone. Although it is likely shorter than that, we believe that this is close to the physical limits due to the inherent mechanical characteristics of the process. We measured the conversion efficiencies our sample to be $3.3\times10^{-4}$, $6.9\times10^{-4}$, and $6.0\times10^{-4}$ in the GHz, THz and optical regimes (1.5 μm), respectively. This is 41 times larger than that of the state-of-the-art MWCNT ($1.7\times10^{-5}$) light-sound converter [19, 20].

As a first application towards a useful device, we employed the properties of the 3G-sponge for the realization of an audio receiver based on an amplitude demodulation at microwave frequencies. The standard communication setup used for this experiment is shown in Fig. 3A and is described in details the Materials and Methods section. Surprisingly, the 3G-sponge was able to perform the demodulation process and generate audible sound. For a demodulator, it is useful to show the characteristic current–voltage (I-V) characteristic (Fig. 3B). The I-V for the 3G-sponge shows a perfect symmetry around the zero crossing point. This characteristic is advantageous over the conventional electronic diodes where the I-V diagram is highly asymmetric, requiring a bias voltage to be applied. Second, the I-V correlation shows a linear component for lower voltages and a cubic term for higher voltages while the power-voltage characteristic (Fig 3B lower plot) is quadratic. In the experiment the modulated signals create thermal oscillations at the 3G-sponge-air interface, which are then converted to sound waves. The demodulated signal (acoustic wave) successfully preserves the quality and information of the modulating one. Due to the symmetric even power-voltage characteristic combined with the demodulation scheme, the 3G-sponge-based demodulator can effectively suppress odd order intermodulation. This leads to very high fidelity in the demodulation since spurious components are suppressed. Figure 3C demonstrates that the original signal $V_{signal}(t)$ and the demodulated acoustic one $V_{sound}(t)$ present identical spectral characteristics. It is worth noting that the presented innovative concept can be applied for the realization of universal and simple demodulation and detection devices. The results of the experiments performed both at optical and THz/microwave frequencies show that this demodulation scheme is independent of the frequency of the carrier. Moreover it can be applied to high frequencies without dedicated advanced electronics. This makes the 3G-sponge a promising material for next generation graphene electronics.

In conclusion our results show a direct and efficient (nearly 41 times larger efficiency than present photo-thermo-acoustic converters) universal coupling between the photonic and phononic spectra in a novel 3D graphene sponge which offers unique elasticity and compressibility. This makes the graphene sponge a very good EM absorber and broadband detector from microwave to optical frequencies. The remarkable coupling in 3G-sponge between light, heat and sound is well-suited for remote heat and sound emission by light. Such properties allowed us to demonstrate a novel heat-mediated frequency demodulator scheme in the microwave range. These unique properties open new opportunities for next-generation graphene electronics and trigger novel research activities for EM wave detection, as well as efficient sound and heat devices controlled by an amplitude-modulated light beam.

**Materials and Methods:**

THz source: Terahertz pulses at 100 Hz repetition rate were produced by optical rectification of an intense mid-infrared femtosecond laser source in an organic crystal. See ref. [12,13] for further details. The THz low pass filters are multi mesh and commercially available.

Setup of the standard communication system (Fig. 3A): We set up a standard communication system where the lower frequency signal $s(t)$, corresponding to the information to be transmitted, is carried on a higher frequency carrier $f_0$ allowing for long distance propagation. The carrier is also modulated in amplitude with a square wave $s(t)$ at 40 Hz with a modulation index of 70% and a duty cycle of 25%. In our experiment, we adapted a demodulation scheme where we superimposed another high frequency carrier of slightly different frequencies $f_0 + \Delta f$ to the carrier at $f_0$. $\Delta f$ is in the useful part of the frequency spectrum for the high-frequency to phononic conversion. The schematic of the transmitter and receiver circuits used for producing the amplitude modulation (AM) and the demodulation are shown in Fig. 3A. All the devices are connected with coaxial cables and the sample of the 3G-sponge was placed in the middle of an N-type connector. The position of the 3G-sponge was adjusted to assure an adequate matching of the forward microwave and to guarantee nearly complete absorption of the EM energy.

Transmitter oscillator and analog AM modulator: Rohde&Schwarz SMA 100 A, CW signals with a frequency range from 9 kHz to 6 GHz. We used the frequency of 1 GHz for the carrier $f_0$.

Function generator: TTi TG1010A. We used it to generate the square wave s(t) at 40 Hz with a modulation index of 70% and a duty cycle of 25%.

Local oscillator: Rohde&Schwarz SMA 100 A, frequency range from 9kHz to 6GHz. We used the frequency of 1 GHz to demodulate the carrier $f_0$ and a $\Delta f$ in the range between 0.1 kHz to 100 kHz.

Graphene sponge production: The graphene 3D sponge was fabricated by an *in situ* solvothermal process using ethanol solvent for low concentrations of Graphene Oxide (GO) sheets (~20 to 50 μm lateral dimension). The low concentration GO ethanol solution (0.20–5.00 mg ml$^{-1}$) was then solvothermally treated in a Teflon-lined autoclave at 180 °C for 12 h to form an intermediate solid, having about 1/3 to 1/2 of the former GO solution volume. After the solvothermal reaction, the ethanol-filled sponge was removed from autoclave and completely immersed in a mixture of acetone and ethanol (1:1 in volume). Then water was added to the system in order to substitute the ethanol with water. This step was then repeated about 15–20 times taking ~6 h for each cycle and up to 5–7 days for the whole process. After the solvent exchange process, the water-filled sponge was freeze-dried to remove the remaining water absorbed. Finally, the sample was annealed at 400 °C for an hour in argon to obtain the final 3D graphene sponge as measured in the photoacoustic experiments.

Thermal camera is a FLIR T420. The camera provides resolution of 320x240 pixel. Its accuracy is ±2% or 2°C -IR –The thermal Sensitivity <0.045°C over the temperature range between -20°C and 150°C.

The microphone is an MK202 from Microtech Gefell and is calibrated in frequency from 20 Hz to 40 kHz. The microphone is mounted on the MV210 preamplifier from the same company (Microtech Gefell). The signal is digitalized in a PXI National Instruments acquisition card.

**Acknowledgments:**

We are grateful to Dominique Zehnder and Michael Eichenberger for assistance with the measurements. We thank Sergejs Dementjevs for the loan of the acoustic measuring system.

**Funding:** We acknowledge financial support from the Swiss National Science Foundation (SNSF) (Grant No. 200021_146769 and No. IZLRZ2-164051). MS acknowledges partial funding from the European Community's Seventh Framework Programme (FP7/2007-2013) under grant agreement no. 290605 (PSI-FELLOW/COFUND). CPH acknowledges association to NCCR-MUST.

**Competing interest:** The authors declare no competing interests.

**Data and material availability:** All data needed to evaluate the conclusions in the paper are present in the paper and/or the Supplementary Materials.

**Author Contributions:** The reported phenomenon was accidentally discovered by M.S. and F.G.; M. G. proposed the sponge modulator. M.G., P.C., C.V. and M.S. performed the microwave measurements. THz and optical measurements were done by M.S. and C.V. C.V and P.C. prepared the figures. Y. C. and B. K. prepared the sample. M.S. proposed the paper coupling concept and wrote the draft. S.L. and C.P.H. coordinated the collaboration. All the authors contributed to the discussions and interpretation of the results.

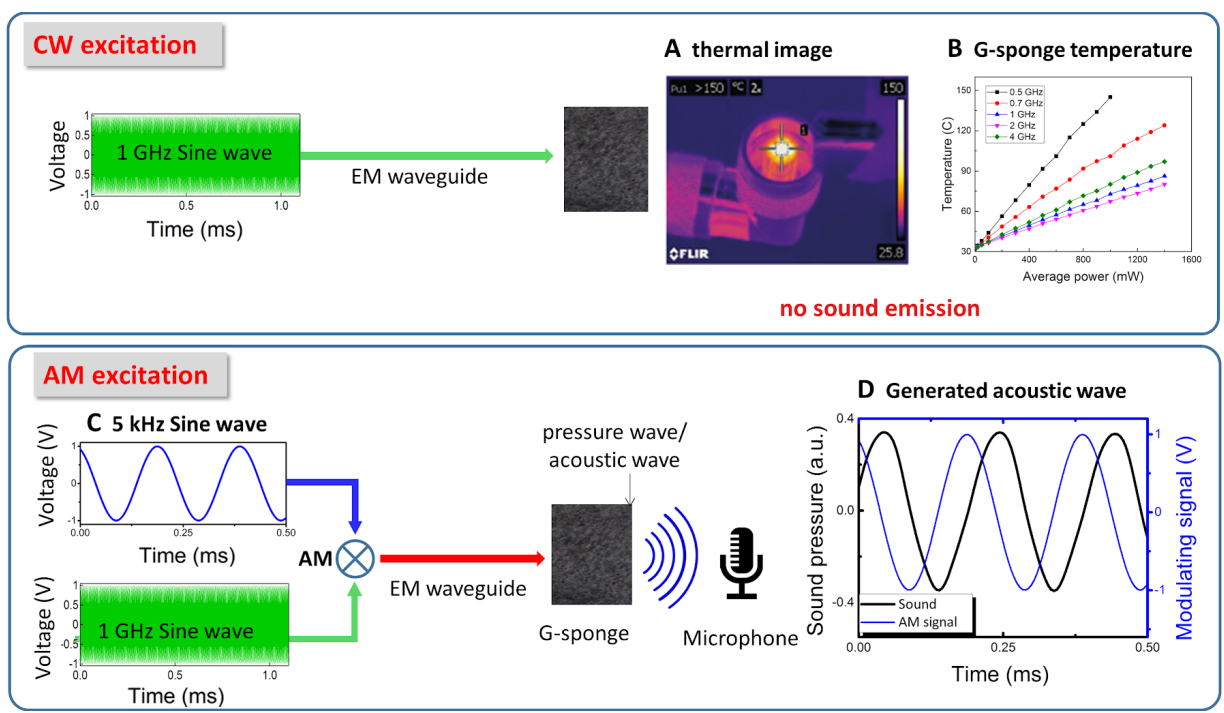

**Fig. 1**. **Photon-phonon coupling in 3D sponge for heat and sound generation.** As electromagnetic excitation we used a microwave sine wave generator at frequency varying between 500 MHz to 4 GHz in CW and amplitude modulation mode. Under the effect of the CW microwave stimulus the graphene sponge heats up rapidly but no sound emission is detectable. The sponge temperature is measured with a thermographic camera (yellow region in (**A**)). The record-high absorption capability and the low thermal inertia of the 3G-sponge leads to rapid and large rise of the temperature beyond 150°C. (**B**) shows the temperature rise as function of the driving microwave power and frequency inside the sample. (**C**) Under the excitation of 1 GHz wave modulated in amplitude by a 5 kHz sine signal, the strong photon-phonon coupling in 3D-sponge results in time varying temperature and pressure wave with consequent sound generation at the modulation frequency show in (**D**). The emitted sound (black) reproduces excellently the modulating signal (blue) and demonstrates the capability of the 3D-sponge for a high-fidelity loudspeaker.

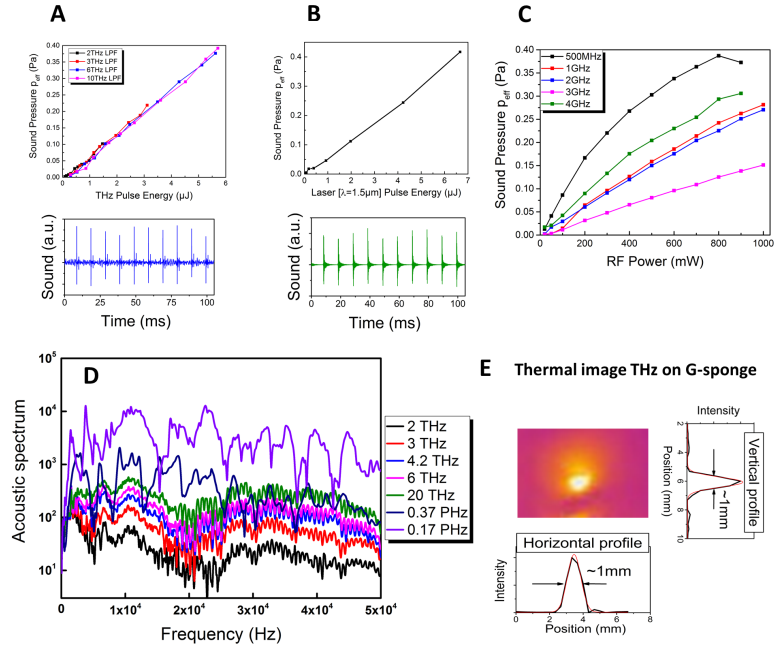

**Fig. 2. Pulsed optical excitation of 3G-sponge in the Terahertz and optical range.** (**A**) The sound amplitude shows a linear dependence on the THz energy at frequency between 1 and 1 THz. Sound emission from 3G-sponge occurs at the repetition rate of the laser (lower graph). Very similar properties are measured at (**B**) optical and (**C**) microwave frequencies. The sound amplitude is linearly dependent on the femtosecond laser pulse energy. (**D**) The recorded sound spectra generated by a pulsed stimulus centered at frequencies across 15 octaves (THz to PHz) are qualitatively comparable. (**E**) The thermal picture of the sponge indicates localized heating at the THz focus. We note that together with a thermal detector the 3G-sponge could thus serve as a 2-dimensional THz intensity profiler.

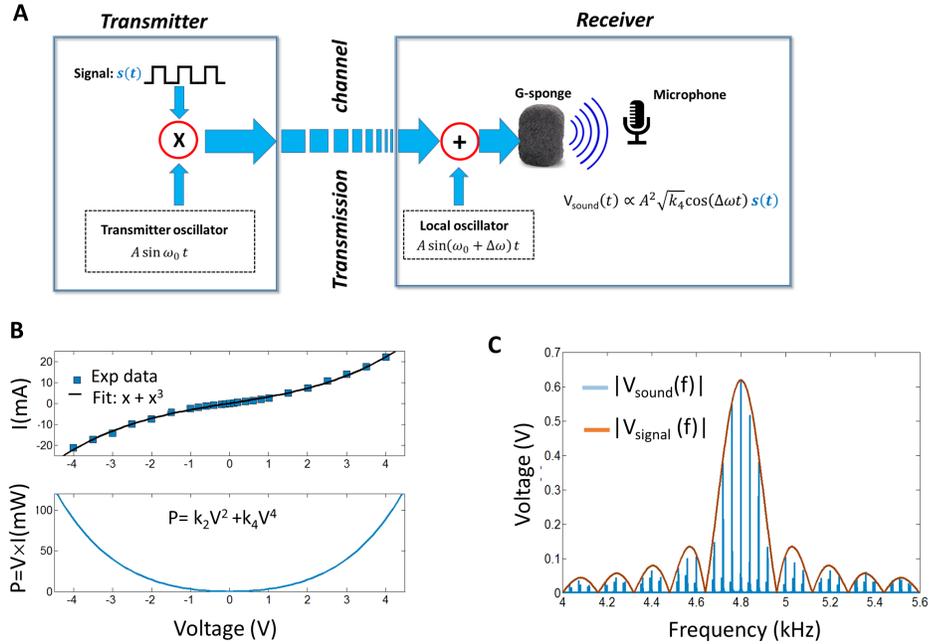

**Fig. 3. Application of 3G sponge as a demodulator using a GHz carrier.** (**A**) Experimental transmitter/receiver communication setup based on the 3G-sponge demodulator. (**B**) Intensity voltage (I–V) characteristics of 3G-sponge. A fit with a pure cubic polynomial is also shown in the plot and corroborates the outstanding symmetry of the 3G-sponge demodulator. The corresponding power-voltage (P-V) characteristic of 3G-sponge is also shown. The nonlinear characteristics usually show a very significant quartic component which makes 3G-sponge an excellent RF demodulator. (**C**) Output spectrum of the acoustic signal $V_{sound}(f)$ measured from the 3G-sponge after its demodulation and carrier suppression corresponds to the transmitted signal spectrum $V_{signal}(f)$.